\documentclass[10pt,twocolumn]{IEEEtran}
\usepackage{graphicx}          
\usepackage{amsmath}
\usepackage{tabularx}
\usepackage{amsfonts}
\usepackage{cite}
\usepackage{url}
\usepackage{verbatim}
\usepackage{booktabs}
\usepackage{times,epsfig}
\usepackage{makecell}
\usepackage{graphicx}
\usepackage[normalem]{ulem}
\usepackage{multirow}
\usepackage{graphicx}
\usepackage[normalem]{ulem}
\useunder{\uline}{\ul}{}
\usepackage{psfrag}
\usepackage{subfigure}
\usepackage{stfloats}
\usepackage{setspace}
\usepackage{amssymb}
\usepackage{multirow}
\usepackage{colortbl}
\usepackage[justification=centering]{caption}
\usepackage{fancyhdr}
\usepackage[table,xcdraw]{xcolor}

\usepackage{enumitem}
\setlist{
    labelindent=\parindent, 
    leftmargin=*,
    nosep
}

\begin{document}
	
	\pagenumbering{arabic}
	
	\title{Generative AI-enabled Mobile Tactical Multimedia Networks: Distribution, Generation, and Perception
	\author{Minrui Xu, Dusit Niyato, Jiawen Kang, Zehui Xiong, Song Guo, Yuguang Fang, and Dong In Kim}
 \thanks{M.~Xu and D.~Niyato are with Nanyang Technological University, Singapore 639798, Singapore. J.~Kang is with Guangdong University of Technology, Guangzhou 510006, China. Z.~Xiong is with the Singapore University of Technology and Design, Singapore 487372, Singapore. S.~Guo is with The Hong Kong Polytechnic University, Hong Kong, China. Y.~Fang is with the City University of Hong Kong, Kowloon, Hong Kong, China. D.~I.~Kim is with Sungkyunkwan University, Suwon 16419, South Korea.}
}
	\maketitle
	\pagestyle{headings}

	\begin{abstract}
Mobile multimedia networks (MMNs) demonstrate great potential in delivering low-latency and high-quality entertainment and tactical applications, such as short-video sharing, online conferencing, and battlefield surveillance. For instance, in tactical surveillance of battlefields, scalability and sustainability are indispensable for maintaining large-scale military multimedia applications in MMNs.
Therefore, many data-driven networking solutions are leveraged to optimize streaming strategies based on real-time traffic analysis and resource monitoring. In addition, generative AI (GAI) can not only increase the efficiency of existing data-driven solutions through data augmentation but also develop potential capabilities for MMNs, including AI-generated content (AIGC) and AI-aided perception. In this article, we propose the framework of GAI-enabled MMNs that leverage the capabilities of GAI in data and content synthesis to distribute high-quality and immersive interactive content in wireless networks. Specifically, we outline the framework of GAI-enabled MMNs and then introduce its three main features, including distribution, generation, and perception. Furthermore, we propose a second-score auction mechanism for allocating network resources by considering GAI model values and other metrics jointly. The experimental results show that the proposed auction mechanism can effectively increase social welfare by allocating resources and models with the highest user satisfaction.
	\end{abstract}

	
\section{Introduction}

Mobile multimedia networks (MMNs) \cite{fujihashi2021soft} have become a critical component in 5/6G, supporting a wide range of applications such as video streaming, video conferences, tactical surveillance, and Metaverse. To make efficient utilization of limited resources in MMNs, AI is leveraged in various ways to determine the streaming configuration of MMNs, including streaming bitrates, buffering strategies, and power allocations for content distribution. {\color{black}Specifically, VR environments can simulate real-world battle scenarios for training. However, VR devices have limited computational capacities and battery life. Therefore, AI-based distribution solutions can extend the virtual operating time and enhance the quality of training.} Moreover, generative AI (GAI) can not only synthesize training datasets for data-driven solutions but also create high-quality and naturalistic AI-generated content (AIGC) in various modalities, e.g., images, videos, and 3D content \cite{bond2021deep}. {\color{black}For example, GAI models can create tactical plans for different scenarios and improve user perception through interactive maps and AI-generated 3D terrain models.}

For data-driven networking solutions in MMNs, GAI can improve delivery efficiency and scalability of multimedia delivery while satisfying user experience by determining network configurations, including streaming bitrate, caching strategies, and transmit power, based on real-time traffic analysis and resource monitoring for tactical surveillance systems~\cite{fujihashi2021soft}. Due to the heterogeneity of user devices in MMNs, optimizing streaming bitrate with flexible GAI models ensures that users experience high-quality and fair multimedia content without interruptions, adapting to their computing capacity and connectivity. As the content requested by users becomes more dynamic, such as short video sharing in tactical surveillance systems~\cite{zhang2022measurement}, GAI-empowered data-driven caching strategies can both enhance user experience by reducing content loading times and decrease network traffic by minimizing real-time data transfers. Finally, adapting transmit power via GAI not only guarantees a stable and longer connection for multimedia streaming but also promotes energy efficiency. 

For intelligent content generation in MMNs, GAI can develop novel content generation~\cite{aldausari2022video}, synchronization~\cite{li2023reparo}, and animation~\cite{ardhianto2023generative} techniques to enrich user experiences without increasing data traffic in GAI-enabled MMNs. By leveraging computing resources, users can leverage GAI models to create, repair, and edit multimedia content~\cite{wang2023videomae, hu2023videocontrolnet, zhou2022cadm}, such as generating music, producing images or video clips from brief descriptions, filling in missing frames in a video, or applying artistic styles to photo beautification directly with their mobile devices, which ensures faster content delivery but also satisfies users' expectations or context. Regarding the received content, users in tactical surveillance systems can adopt super-resolution and interpolation techniques to leverage GAI to enhance the quality of low-resolution images or videos by generating high-definition versions~\cite{lee2021deep} and estimating intermediate frames in videos~\cite{reda2022film}, respectively. Finally, GAI can synchronize different media types and revolutionize animation in multimedia content production, including character design, motion capture, facial animation, scenery and background generation, and interactive animation.

Although GAI offers many advantages for various multimedia applications, there are also many remaining potential challenges to overcome to deploy GAI in MMNs. For resource-constrained mobile devices, long processing times and high energy consumption when running GAI models can affect the quality of service (QoS) and quality of experience (QoE)~\cite{yang2023insights}. Additionally, collecting large amounts of data for pre-training tasks such as super-resolving images and interpolating videos is resource-intensive, and the data may not always match real-world scenarios due to network traffic fluctuations and possible data corruption~\cite{bond2021deep}. Finally, for GAI models used in MMNs, ensuring the stability in the scale of multimedia applications and consistency of generated content is critical to user experiences~\cite{khan2023taxonomy}.

In this article, we propose a framework of Generative AI-enabled MMNs to provide real-time and low-latency multimedia services to wireless users. Under this framework, GAI techniques can not only enhance the efficiency and scalability of content distribution in existing multimedia networks and tactical surveillance systems but also introduce new capabilities for dynamic content generation and perception. Additionally, to allocate network resources to execute generative model-based multimedia services, we propose a second-score auction to optimize the efficiency of resource utilization of edge servers. Compared to the second-price auction, the proposed mechanism can achieve higher revenue improvement when the model scales increase.

\section{The Framework of Generative AI-enabled Mobile Multimedia Networks}

\begin{figure*}
    \centering
    \includegraphics[width=0.9\linewidth]{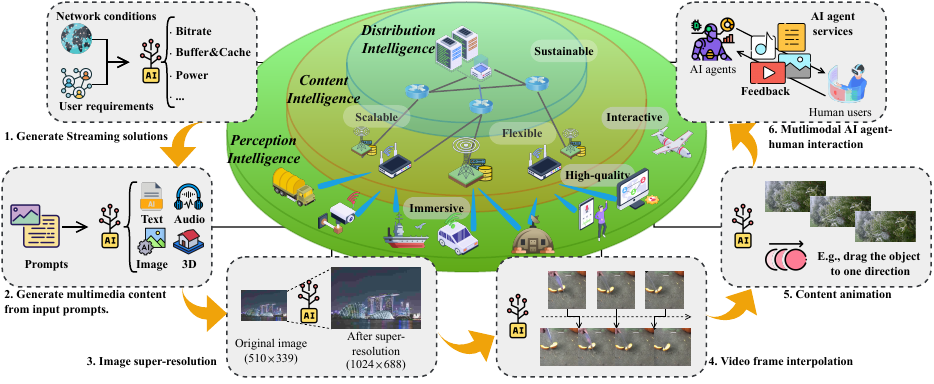}
    \caption{The overview of Generative AI-enabled MMNs, including content distribution solutions, dynamic content generation, quality enhancement, and interactive perception based on GAI techniques.}
    \label{fig:system}
\end{figure*}

Mobile multimedia networks can support the generation, distribution, and perception of multimedia content over mobile and wireless communication systems. In addition to the focus of traditional MMNs on providing static multimedia services to satisfy users, Generative AI-enabled MMNs leverage GAI in network management and content generation for multimedia services to provide dynamic multimedia services.

\useunder{\uline}{\ul}{}

\begin{table*}[t]
\small\centering
{\color{black}
\caption{A comparison of different generative AI algorithms in mobile multimedia networks.}
\label{tab:gai}
\begin{tabular}{|p{.08\textwidth}<{\centering}|p{.2\textwidth}<{\raggedright}|p{.2\textwidth}<{\raggedright}|p{.2\textwidth}<{\raggedright}|p{.2\textwidth}<{\raggedright}|}
\hline
Algorithms & Generative Adversarial Networks & Variational Autoencoders & Diffusion Models & Generative Transformers \\ \hline
Exemplary Applications & \textbf{Image super-resolution}: \begin{itemize}[labelindent=0em, leftmargin=*] \item a generator to upscale low-resolution images \item a discriminator differentiates between the upscaled and original images \item Latency: medium \item Resource cost: medium \item Metrics: PSNR, SSIM \end{itemize} & \textbf{Neural compression}: \begin{itemize} \item learning a compact latent representation of frames \item reconstructing original data with minimal loss of information \item Latency: low \item Resource cost: low \item Metrics: compression ratio,   reconstruction quality \end{itemize} & \textbf{Content animation}: \begin{itemize} \item adding noise iteratively \item denoising via simulating a diffusion process to generate dynamic visual transformations \item Latency: high \item Resource cost: high \item Metrics: temporal coherence, user acceptance \end{itemize} &  \textbf{Cross-modal transferring}: \begin{itemize} \item extracting features using self-attention \item generating coherent responses by aligning input and output modalities \item Latency: low \item Resource cost: high \item Metrics: BLEU, GLUE \end{itemize} \\ \hline
Advantages & \begin{itemize} \item High-resolution and realistic image generation \item Versatile models for novel content manipulation \item Fast and accurate dataset augmentation \end{itemize} & \begin{itemize} \item Consistency as continuous and structured latent \item Enhanced scalability via compression \item Suitable compression for limited system resources \end{itemize} & \begin{itemize} \item Applicable for various applications requiring flexible qualities \item Flexible generative behavior during the configuration of steps \end{itemize} & \begin{itemize} \item Capture long-range dependencies of input data \item Trained for cross-modal transferring tasks \item Design interactive tools for multimodal services \end{itemize} \\ \hline
Limitations & \begin{itemize} \item Potential mode collapse, e.g., vanishing gradients \item Unable to generate simulation data with the same intrinsic dimension \item Risky for healthcare \end{itemize} & \begin{itemize} \item Lossy data compression causes quality loss \item Inefficient approaches for the separate design of applications and networks \item Pre-defined adaptability \end{itemize} & \begin{itemize} \item Require multiple steps for content generation, introducing latency \item Not ideal for real-time streaming applications on mobile devices \end{itemize} & \begin{itemize} \item Memory-intensive during dealing with long sequences content \item Perpetuate or even amplify biases present in the training data \end{itemize} \\ \hline

\end{tabular}%
}
\end{table*}

\subsection{Generative AI for Multimedia Networks}

GAI is a promising area for synthesizing multimodal multimedia content, including text, images, videos, and point clouds. As listed in TABLE~\ref{tab:gai}, there are several advanced GAI models, e.g., Generative Adversarial Networks (GANs), Variational Autoencoders (VAEs), flow-based models, and diffusion models\cite{bond2021deep}. GANs first train the generator and the discriminator through adversarial training to achieve equilibrium between them. At the end of the training phase, the discriminator cannot distinguish between the real data and the data generated by the generator~\cite{aldausari2022video}. In addition, VAEs use an encoder to compress the input data into a latent space that generates a probability distribution for each variable, and a decoder to reconstruct the data from this space using a combination of reconstruction loss and regularization. Furthermore, diffusion models iteratively add noise to the data and refine and denoise it to produce a sample that resembles the target data distribution.

The use of GAI in MMNs can bring various benefits, from content delivery to generation~\cite{khan2023taxonomy}. In content delivery of tactical surveillance systems, GAI can help operators ensure QoS in MMNs by providing insights and making intelligent decisions on streaming and generating of multimedia content. Moreover, GAI can improve existing data-driven methods in MMNs by enriching their training datasets for more accurate and robust performance. Finally, GAI models can generate high-quality multimedia content in a plug-and-play manner to improve the user experience for multimedia applications, e.g., tactical surveillance, without training and fine-tuning.

\subsection{Multimedia Distribution Intelligence}

\subsubsection{Dynamic Adaptive Streaming}

Dynamic adaptive streaming is an efficient method of delivering content over the mobile Internet that adapts to changing network conditions and device capabilities. This is achieved by dividing a video into small segments and encoding them at different quality levels. The mobile device then selects an appropriate segment based on current conditions, such as available bandwidth, and seamlessly switches between quality levels as needed. Fortunately, GAI techniques, such as super-resolution (SR) neural models, can be used to improve the quality of video streams in dynamic adaptive streaming~\cite{lee2021deep}. These models can increase the resolution of lower-quality video frames to improve the visual quality of the stream. For example, diffusion models with their ability to generate images can be used to improve the tradeoff between rate and distortion in dynamic adaptive streaming~\cite{zhou2022cadm}. 
By integrating GAI into dynamic adaptive streaming, significant bitrate reduction can be achieved while maintaining high restoration quality. This can lead to better video transmission performance and a better streaming experience for users.

\subsubsection{Buffering and Caching}

Buffering and caching are techniques used in multimedia networks to improve user experience and reduce network load. In buffering, multimedia data is stored to ensure smooth playback despite network delays and fluctuations in data transmission. Caching stores frequently accessed multimedia content near a user, reducing repeated data transfers from the source. {\color{black}By predicting user preference in MMNs, intelligent buffering and caching algorithms based on GAN-empowered deep reinforcement learning algorithms can accelerate convergence by synthesizing the augmented dataset and thus minimize the overall transmission cost in MMNs where popularity fluctuates over time~\cite{weng2020content}.} Therefore, GAI-enabled buffering and caching strategies can help reduce core network traffic and transmission energy consumption, improving overall network efficiency.

\subsubsection{Energy Efficiency}

In MMNs, energy efficiency is critical due to the limited battery budget in mobile devices, like drones. {\color{black}GAI can help improve the quality of compressed video chunks at the receiver by using super-resolution techniques. For instance, GANs can dynamically select the network architecture used for super-resolution, allowing a trade-off between image quality and computational complexity~\cite{choi2021delay}. By adjusting the network depth,} GANs can balance image quality with the CPU clock frequencies required for operation, leading to higher energy efficiency. To achieve energy efficiency in MMNs, it is important to adaptively compress video chunks, dynamically adjust the depth of the neural network, and optimize it to minimize computational overhead.

\subsection{Multimedia Content Intelligence}

{\color{black}As the evolutionary feature compared to traditional CDNs, GAI-enabled MMNs can generate intelligent content, which is dynamic and personalized, during real-time delivery.}

\subsubsection{Content Synthesis, Repairing and Editing}

GAI technology can be used to create, repair, and edit multimedia content in real-time based on user preferences or context. For example, a music application could generate a unique piece of music based on a user's current mood or activity. In addition, a user can enter a short description, and an application uses a GAI model to generate a corresponding image or short video clip. When streaming over MMNs, if certain frames of a video are missing or corrupted, a GAI model can predict and fill in the missing frames to ensure an uninterrupted viewing experience. When editing content, users can use GAI to change the weather in a photo (e.g., turn a sunny scene into a rainy one) or adjust the instruments in a music track. In style transfer, the artistic style of one image can be applied to another~\cite{hu2023videocontrolnet}. For example, a user can take a photo and apply the style of a famous painting directly to their mobile devices or via server-side processing.

\subsubsection{Image Super-resolution and Video Interpolation}

To improve perceptual quality on the client side, super-resolution of images and video interpolation are useful techniques that GAI employs to improve the quality of received content before rendering~\cite{lee2021deep}. Super-resolution is a technique that aims to improve the resolution and quality of low-resolution images or videos by generating high-resolution versions, while interpolation is a technique used to estimate intermediate frames in videos, allowing systems to increase the achieved frame rate and reduce friction losses. GAI based on Convolutional Neural Networks (CNNs) is used for super-resolution tasks in content delivery systems that can generate high-resolution images from low-resolution images to improve visual quality. 



\subsubsection{Cross-modal Content Synchronization and Animation}

GAI enables the automatic synchronization of multimedia content from different modalities. For example, a GAI model trained in speech recognition can create and synchronize text transcriptions for videos without subtitles in real-time. Similarly, a GAI model can create corresponding audio annotations and synchronize them with visual content for text descriptions. In addition, animation is the process of creating moving objects from 2D images to 3D point clouds or other visual representations. In the meanwhile, GAI can revolutionize the animation process, including character design, motion capture, facial animation, scenery and background generation, and interactive animation. {\color{black}For example, in command and control systems, visual aids or diagrams can appear in sync with the narration while a character explains the current status~\cite{ardhianto2023generative}. By leveraging GAI in motion capture, realistic facial expressions and movements can be synthesized to help the interpretation of the static data.} 

\subsection{Multimedia Perception Intelligence}

\subsubsection{Personalized Content Translation and Style-Transfer}

{\color{black}After receiving multimedia content, GAI algorithms based on neural networks can translate and modify the style of the content with low computation complexities to enhance the personalized experience for users.} By using GAI techniques, content translation can make it accessible and comprehensible to a wide range of audiences, e.g., users in international operations. GAI-enabled MMNs can identify the language and preferences of a user receiving the content and can present the original content in a user's preferred language through real-time language translation. Additionally, GAI-enabled MMNs can replace inappropriate segments of original content with more suitable segments from the target content in real-time, based on the rating given by a viewer of tactical surveillance systems. For instance, GAI can apply style transfer to transform graphic and violent images in an original video into a cartoon character's fight, making it more suitable for children, depending on a viewer's age.

\subsubsection{Generative AI Agent-Human Multimodal Interaction}

Based on the perception, users in GAI-enabled MMNs can interact with GAI agents for continuous improvement of the QoE. In multiple and deep interactions, users can leave more user profiles at GAI agents to enjoy higher quality and more consistent multimedia services in the future. During the process of interacting with GAI agents, users can use natural language to send requests to the agents and provide feedback on the perceived content and service quality. GAI agents with memory capability can save the past interaction data into the model and use it as a basis for future service delivery. As a result, GAI agents can serve as personal assistants who understand customer queries in different formats and provide detailed, multimodal responses. For instance, they can employ planning methods such as chain-of-thought and tree-of-thought to think deeply before offering AI services. In educational multimedia applications, AI can generate customized learning materials that incorporate text, images, and audio to accommodate different learning styles. However, the interaction process of GAI agents typically requires more resources compared to static GAI-based multimedia services.

\section{Potential Applications and Challenges in Generative AI-enabled MMNs}

\subsection{Potential Applications}

\subsubsection{Short Video Sharing}

Short video-sharing applications are for sharing videos that are only a few seconds to a few minutes long on social media platforms such as TikTok, Douyin, and Kuaishou. Compared to traditional video-sharing platforms, short video sharing has unique characteristics, such as shorter video duration, different user behavior, and different popularity dynamics. As analyzed in~\cite{zhang2022measurement}, short video-sharing services are conservative in bitrate selection and require real-time caching strategies because the streaming duration is relatively short, and the popularity changes rapidly. Moreover, GAI can provide short video services by synthesizing natural videos on the fly based on usage behaviors and patterns. 



\subsubsection{Metaverse}

Metaverse is a collective 3D virtual shared space that users can access via augmented/virtual reality (AR /VR). To support seamless interaction between the physical world and virtual spaces, a huge amount of visual and haptic data must be transmitted for real-time synchronization in building the virtual spaces from comprehensive sensor data, while simultaneously providing multimodal feedback back to the physical world. To reduce the size of the transmitted data without losing fidelity, VAEs can perform neural compression, which encodes the content into a compressed representation and then decodes them back to their original form~\cite{yang2023insights}. 


\subsection{Potential Challenges}

\subsubsection{Computational Complexity} The execution of GAI models for short video creation and real-time multimodal animation is computationally intensive, which can pose a challenge for real-time performance on mobile devices. This results in longer processing times and higher energy consumption which hinders users' interest in using GAI-based applications. To address this challenge, optimizing the model architecture and inference process is also possible to reduce the computational complexity for real-time applications. For example, the use of edge computing and distributed learning to run GAI-based tactical surveillance systems can help reduce the computational overhead of devices and play the video in time by offloading computation tasks to edge servers. 

\subsubsection{Training Data Collection} Aside from the computational and communication overhead, it is also difficult to meet the requirements for large amounts of pre-training data to produce good results for image super-resolution and video interpolation tasks. In MMNs with varying streaming conditions, such as different network bandwidths, device capabilities, and user preferences, collecting such datasets for pre-training of GAI models can be time and resource-intensive, and it can be difficult to ensure that they are representative of real-world scenarios. In addition, data in MMNs may be incomplete or corrupted due to factors such as network congestion and packet loss, further complicating the training process. 


\section{Use Case: Multi-dimensional Auction for Resource Allocation of Multimedia GAI models}

\begin{figure*}
    \centering
    \includegraphics[width=1\linewidth]{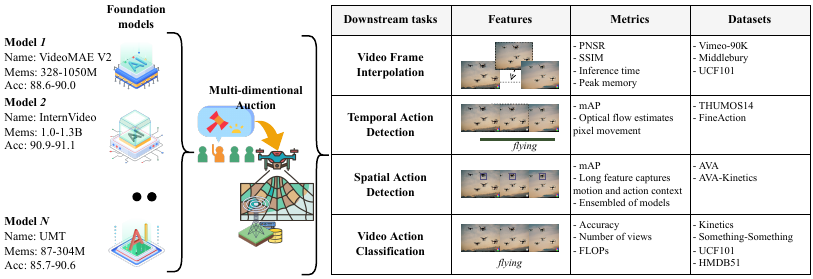}
    \caption{Model markets for Generative AI-enabled MMNs with multiple GAI models, e.g., VideoMAE V2 (https://github.com/OpenGVLab/VideoMAEv2), InternVideo (https://github.com/OpenGVLab/InternVideo), UMT (https://github.com/TencentARC/UMT), for various downstream tasks.}
    \label{fig:models}
\end{figure*}


\begin{figure}
    \centering
    \includegraphics[width=1\linewidth]{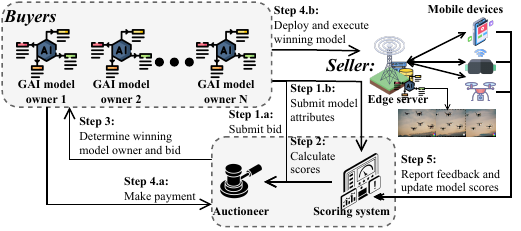}
    \caption{The workflow of proposed auction-based resource allocation for GAI models in Generative AI-enabled MMNs.}
    \label{fig:workflow}
\end{figure}


GAI-enabled MMNs can deploy GAI models on edge servers for provisioning real-time, high-quality multimedia services{\color{black}, such as tactical planning, and soldier training, crisis management.} During the simulation and rendering of battlefield environments, the execution of these GAI models consumes a large amount of memory, computation, and communication resources of edge servers. Furthermore, these GAI models may have multiple attributes for evaluating services {\color{black}during the operation of tactical surveillance,} such as image super-resolution, video interpolation, and content animation. Therefore, allocating resources to edge servers to maximize the utility of MMNs is not trivial.

In this use case, we consider an auction-based resource allocation for {\color{black}GAI-enabled tactical surveillance applications. In the market of GAI models for tactical surveillance,} bidders are owners of GAI models, and they would like to compete for limited resources in edge networks. Edge servers, which have storage, computation, and communication resources, act as sellers and auctioneers. To determine which models to load and execute to provide multimedia services in the current time window, an auctioneer must determine the winners based on the information it receives from bidders. Finally, the auctioneer helps transfer the monetary revenue to the sellers, i.e., the edge servers. However, existing auction mechanisms, such as second-price auctions, cannot fully capture the information needed to determine the allocation results, resulting in a loss of market efficiency. Fortunately, bidders can also report their model attributes in addition to the bidding price to the auctioneer for accurate evaluation of the score of their bids. Finally, by developing a scoring system for these models, the auctioneer leverages second-score auctions to determine the winning bidder and the corresponding payment.


As illustrated in Fig.~\ref{fig:workflow}, the workflow of the second-score mechanism consists of five main steps as follows.

\textbf{Step 1:} At the beginning of the auction, buyers, i.e., GAI model owners submit their multidimensional bids, including bidding price and other model attributes, including latency, resource cost, and performance, to the auctioneer.

\textbf{Step 2:} Based on the received model attributes, the auctioneer calculates the scores as the weighted sum of basic value (for the model) and execution values (for the normalized performance of running tasks required by users) of these bids based on the scoring system.

\textbf{Step 3:} According to the score of each bid, the auctioneer determines the winner as the bidder of the bid with the highest score and the corresponding payment as the bidding price of the bidder of the bid with the second highest score

\textbf{Step 4:} The winning GAI model owner makes payment to the edge server and deploys the model on the edge server for providing multimedia services to GAI-enabled MMNs.

\textbf{Step 5:} Based on the user feedback reported from MMNs, the auctioneer updates the scoring system, which consists of all historical reports of GAI models.

{\color{black}As illustrated in Fig. 3, after the completion of content delivery, the users report their feedback and update model scores for the contents obtained from the GAI models based on both subjective evaluation, i.e., user acceptance, and objective evaluations, i.e., content quality.}

\begin{figure}
    \centering
    \includegraphics[width=0.8\linewidth]{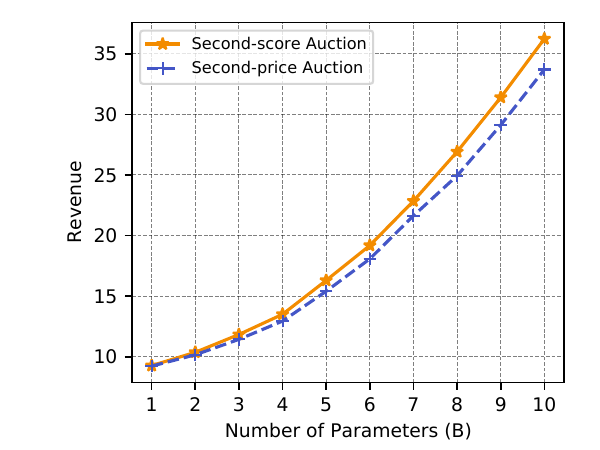}
    \caption{The market revenue versus the size of foundational models for mobile multimedia services.}
    \label{fig:auction}
\end{figure}

{\color{black} The auction mechanisms optimize the efficiency of resource utilization of edge servers in GAI-enabled MMNs, leading to improved revenue.} Compared to the second-price auction, the second-score auction can incorporate multiple price irrelevant attributes, e.g., latency and accuracy, during the allocation and pricing phases. In this way, the total social welfare can be improved under well-established scoring systems. As illustrated in Fig~\ref{fig:models}, we perform the simulation experiment to evaluate the performance of the second-score auction mechanism with 10 model owners in the network system. The basic value of each model ranges from 0 to 10 and the executing value of each task ranges from 0 to 1. Based on the scaling law of GAI models, we consider the number of downstream tasks of GAI models that can be tackled as the square of the model size (in billions). By ranging the number of parameters of models from 1 billion to 10 billion, we obtain the results shown in Fig.~\ref{fig:auction}. By considering the additional model attributes, the second-score auction can achieve high revenue for edge servers and improve the utilization of resources in MMNs.

{\color{black}\section{Future Research Directions}

\textbf{Lightweight Generative AI Models:} GAI-enabled MMNs incur additional computation and communication overhead during content generation due to the use of GAI models. For faster inference and low-cost execution, lightweight GAI models with fewer weights and smaller model architecture can be used to reduce training and inference overhead during making distribution decisions or generating battlefield models in military multimedia systems. Once trained on large datasets, GAI models can be optimized for low requirements through pruning, quantization, and knowledge distillation. 

\textbf{AI Model Caching:} In the last decades, content caching for CDNs has attracted tremendous attention from both academia and industry. However, in MMNs, content, such as animations, sound effects, and interactive scenarios on battlefields, is generated by GAI models which can be diverse and dynamic according to the input of GAI models. Therefore, content caching is less effective as new content can be synthesized anytime and anywhere by executing the GAI models. In this regard, caching for GAI models is introduced as storing and reusing GAI models at edge servers that reduce latency and resource consumption. 

\textbf{Privacy-preserving Content Generation:} GAI can provide a more efficient framework for the distribution, generation, and perception of multimedia content that protects privacy while maintaining the desirable features of streaming, synthesizing, and analytics systems for multimedia content. 
Specifically, GAI can protect privacy by enabling more privacy-preserving video transformation during the recreation of real-world battle scenarios for training purposes. 
}
\section{Conclusions}

In this article, we have proposed a framework of generative AI-enabled MMNs for scalable, sustainable, and immersive multimedia services. In this framework, we have provided a tutorial on how GAI techniques can improve the efficiency of data-driven content distribution solutions under existing multimedia network infrastructure and provide novel capabilities in content generation and perception for future immersive multimedia applications. Moreover, we have presented some potential applications in GAI-enabled MMNs and highlighted the potential challenges in this framework. Furthermore, we have proposed a second-score auction mechanism to allocate resources to execute GAI-based multimedia services in GAI-enabled MMNs. 

\bibliographystyle{ieeetr}
\bibliography{main}
	\begin{IEEEbiographynophoto}{Minrui Xu} [S’23] (minrui001@e.ntu.edu.sg)
    received the B.S. degree from Sun Yat-Sen University, Guangzhou, China, in 2021. He is currently working toward a Ph.D. degree in the School of Computer Science and Engineering, at Nanyang Technological University, Singapore. His research interests mainly focus on Metaverse, quantum information technologies, deep reinforcement learning, and mechanism design.
    \end{IEEEbiographynophoto}
    
    \begin{IEEEbiographynophoto}{Dusit Niyato} [M'09, SM'15, F'17] (dniyato@ntu.edu.sg)
        is currently a professor in the School of Computer Science and Engineering, Nanyang Technological University, Singapore. He received a B.Eng. degree from King Mongkuts Institute of Technology Ladkrabang (KMITL), Thailand in 1999 and a Ph.D. in electrical and computer engineering from the University of Manitoba, Canada in 2008. His research interests are in the areas of the Internet of Things (IoT), machine learning, and incentive mechanism design.
    \end{IEEEbiographynophoto}
    \begin{IEEEbiographynophoto}{Jiawen Kang} [M'18]
        (kavinkang@gdut.edu.cn) received the M.S. degree and the Ph.D. degree from the Guangdong University of Technology, China, in 2015 and 2018, respectively. He is currently a full professor at the Guangdong University of Technology. He was a postdoc at Nanyang Technological University from 2018 to 2021, Singapore. His research interests mainly focus on blockchain, security, and privacy protection in wireless communications and networking.
    \end{IEEEbiographynophoto}

    \begin{IEEEbiographynophoto}{Zehui Xiong} [M'20]
        (zehui\_xiong@sutd.edu.sg) is an Assistant Professor at the Singapore University of Technology and Design. Before that, he was a researcher with Alibaba-NTU Joint Research Institute, Singapore. He received a Ph.D. degree in Computer Science and Engineering at Nanyang Technological University, Singapore. He was a visiting scholar at Princeton University and the University of Waterloo. His research interests include wireless communications, network games and economics, blockchain, and edge intelligence.
    \end{IEEEbiographynophoto}
    \begin{IEEEbiographynophoto}{Song Guo}
        (song.guo@polyu.edu.hk) received his Ph.D. degree in Computer Science from University of Ottawa. Now, he is a Full Professor at the Department of Computing at The Hong Kong Polytechnic University. His research interests are big data, cloud computing, mobile computing, and distributed systems.
    \end{IEEEbiographynophoto}
        \begin{IEEEbiographynophoto}{Yuguang Fang}
        (my.fang@cityu.edu.hk) received an M.S. degree from Qufu Normal University, China in 1987, a Ph.D. degree from Case Western Reserve University in 1994, and a Ph.D. degree from Boston University in 1997. He joined the Department of Electrical and Computer Engineering at the University of Florida in 2000 as an assistant professor, then was promoted to associate professor in 2003, full professor in 2005, and distinguished professor in 2019, respectively. Since 2022, he has been the Chair Professor of Internet of Things with the Department of Computer Science at City University of Hong Kong.
    \end{IEEEbiographynophoto}
\begin{IEEEbiographynophoto}{Dong In Kim}
        (dikim@skku.ac.kr) Dong In Kim received his Ph.D. in electrical engineering from the University of Southern California, Los Angeles, CA, USA, in 1990. He is a Professor with the College of Information and Communication Engineering, Sungkyunkwan University, Suwon, South Korea. His research interests include Internet of Things, wireless power transfer, and connected intelligence.
    \end{IEEEbiographynophoto}
\end{document}